\documentclass[a4paper,fleqn]{cas-sc}

\usepackage[authoryear]{natbib}

\def\tsc#1{\csdef{#1}{\textsc{\lowercase{#1}}\xspace}}
\tsc{WGM}
\tsc{QE}

\usepackage[utf8]{inputenc}
\usepackage{booktabs}
\usepackage{graphicx}
\usepackage{multirow}
\usepackage{array}
\usepackage{amsmath}
\usepackage{array}

\newcommand{\PreserveBackslash}[1]{\let\temp=\\#1\let\\=\temp}

\usepackage{newunicodechar}
\newunicodechar{≤}{\ensuremath{\leq}}
\newunicodechar{≥}{\ensuremath{\geq}}

\newcolumntype{C}[1]{>{\PreserveBackslash\centering}p{#1}}
\newcolumntype{R}[1]{>{\PreserveBackslash\raggedleft}p{#1}}
\newcolumntype{L}[1]{>{\PreserveBackslash\raggedright}p{#1}}

\usepackage{tabularx}
\usepackage{lscape}  
\usepackage[figuresright]{rotating}

\begin{document}
\let\WriteBookmarks\relax
\def\floatpagepagefraction{1}
\def\textpagefraction{.001}

\shorttitle{A Meta‑analysis of College Students’ Intention to Use Generative Artificial Intelligence}    

\shortauthors{Diao et al.}  

\title [mode = title]{A Meta‑analysis of College Students’ Intention to Use Generative Artificial Intelligence} 
\medskip{}

%

\vspace{1em}
\author[1]{Yifei Diao}
\author[2]{Ziyi Li}
\author[3]{Jiateng Zhou}
\cormark[1]
\ead{jiatengzhou_edu@163.com}
\author[1]{Wei Gao}
\cormark[1]
\ead{gaowei@ccnu.edu.cn}
\author[1]{Xin Gong}
\cormark[1]
\ead{gongxin@ccnu.edu.cn}

\affiliation[1]{organization={School of Education},
                addressline={Central China Normal University}, 
                city={Wuhan},
                postcode={430079}, 
                country={China}}      

\affiliation[2]{organization={Faculty of Education},
                addressline={University of Cambridge}, 
                city={Cambridge},
                postcode={CB2 8PQ}, 
                country={United Kingdom}}   

\affiliation[3]{organization={Faculty of Education},
                addressline={Beijing Normal University}, 
                city={Beijing},
                postcode={100875}, 
                country={China}}

\cortext[cor1]{Corresponding author}
\cortext[cor2]{These authors contributed equally: Yifei Diao, Ziyi Li, Jiateng Zhou}

\begin{abstract}
\nocite{}
It is of critical importance to analyse the factors influencing college students' intention to use generative artificial intelligence (GenAI) to understand and predict learners' learning behaviours and academic outcomes. Nevertheless, a lack of congruity has been shown in extant research results. This study, therefore, conducted a meta-analysis of 27 empirical studies under an integrated theoretical framework, including 87 effect sizes of independent research and 33,833 sample data. The results revealed that the main variables are strongly correlated with students' behavioural intention to use GenAI. Among them, performance expectancy (r = 0.389) and attitudes (r = 0.576) play particularly critical roles, and effort expectancy and habit are moderated by locational factors. Gender, notably, only moderated attitudes on students' behavioural intention to use GenAI. This study provides valuable insights for addressing the debate regarding students' intention to use GenAI in existed research, improving educational technology, as well as offering support for school decision-makers and educators to apply GenAI in school settings.
\end{abstract}



\maketitle
\section{Introduction}\label{}

Generative Artiﬁcial Intelligence (GenAI), such as ChatGPT, has been increasingly becoming a signiﬁcant impact on contemporary educational development, due to its powerful capabilities of information processing and knowledge production \citep{wu2023brief, kocon2023chatgpt,gilardi2023chatgpt}. GenAI, being even considered as ‘the brightest student in the class’ \citep{vazquez2023chatgpt}, is evidenced to excel in various aspects, such as natural language processing\citep{nazir2023comprehensive}, image analysis \citep{borger2023artificial}, computer programming \citep{wang2023chatgpt}, and artistic creation \citep{guo2023can}. Therefore, an increasing number of students are now seeking GenAI to support their learning processes \citep{jo2023decoding,li2023rethinking,yusuf2024generative}. 

Examining students’ behavioural intentions to use GenAI tends to be crucial, owing to the strong connection between practical behaviours  \citep{liu2024measuring,liu2024qualitative}and academic outcomes \citep{zhai2020understanding}. Although extant studies have explored the factors influencing students’ GenAI behavioural intentions from different perspectives, their findings are inconsistent, resulting in a lack of congruity\citep{von2023artificial}. Furthermore, existing research has mostly been conducted under one-sided theoretical frameworks, lacking multidimensional analyses of the influencing factors. As a result, this study conducts a meta-analysis of 27 empirical studies over the last five years, involving various core variables in different theoretical contexts to assess previous findings and view established debates in existing research comprehensively.

Meta-analysis is defined as “the statistical analysis of a large collection of analysis results from individual studies to integrate the findings”\citep{glass1976primary}, considered to be suitable for analysing studies involving heterogeneity \citep{hunterschmidt}. Notably, no research has examined the factors influencing college students’ behavioural intentions to use GenAI through a meta-analysis approach.

We drew on the insights of Dai et al.\citep{dai2024meta} to formulate our research questions, which included:

RQ1. What are the key theories and core variables mentioned in research on the factors influencing college students’ behavioural intentions to use GenAI?

RQ2. Which of these variables has stronger influences on college students’ behavioural intentions to use GenAI?

RQ3. Can region and gender moderate between these variables and college students’ behavioural intentions to use GenAI?

By exploring the research questions above, this study aims to address the debates in extant research, support building a more comprehensive theoretical perspective, and provide practical insights for multiple subjects, including technology developers, school policymakers, and teachers, particularly under the background of educational change led by GenAI.

\section{Literature Review and Theoretical Framework}\label{}

\subsection{Literature Review}

Early research on GenAI focuses on ChatGPT. ChatGPT is rather so appealing that it has attracted more than one million users in five days, since its release in November 2022\citep{dwivedi2023opinion}. Initially, students typically used ChatGPT out of curiosity \citep{hmoud2024higher,vsedlbauer2024students}. Yet, some students subsequently applied ChatGPT in academic works \citep{zhao2024chatgpt,bavsic2023chatgpt}, and such behaviours carry the potential risk of academic misconduct, which further raises concern for higher education institutions \citep{kasneci2023chatgpt, thorp2023chatgpt}. A large amount of prestigious tertiary education institutions, as well as compulsory education systems, have issued prohibitions to regulate those who may infringe the regulations with the risk of expulsion from school \citep{yu2023reflection}. In addition, information security and ethical issues raised by ChatGPT also have emerged \citep{charfeddine2024chatgpt, okey2023investigating}. Nowadays, the factors influencing students’ behavioural intentions to use GenAI are much more complex than before, influenced by multiple forces. Notably, technology utilisation behaviour is no longer driven only by inner curiosity, but also by students’ perceptions of the technology, as well as external factors such as organisational culture \citep{martinez2020digital} and social norms \citep{kumar2020behavioral}.

Although many scholars have explored the factors influencing students’ utilisation of GenAI, they have not yet reached final conclusion. Firstly, from the effect size perspective, Rahman et al. \citep{rahman2022examining}confirmed that perceived ease of use significantly affects students’ behavioural intention to use ChatGPT, while Yilmaz et al. \citep{yilmaz2023generative} disagreed. In addition, whether trustworthiness has a significant effect remains controversial as well\citep{hernandez2023information,jo2023understanding}. Even in the same study, different research methods can lead to contradictory conclusions, especially in Foroughi et al.’s study\citep{foroughi2023determinants} . They found that factors, such as social influence, facilitating conditions, and habit, do not affect students’ use of ChatGPT by partial least squares regression, but subsequent exploration by fuzzy-set qualitative comparative analysis found conflicting findings. Secondly, in terms of the influence mechanism, Zou and Huang  \citep{zou2023use}found that perceived usefulness directly influences students’ intention to use ChatGPT, but Maheshwari \citep{maheshwari2023factors} pointed out that perceived usefulness indirectly impacts students through personalisation and interactivity. In sum, a meta-analysis of existing research to figure out these conflicted arguments tends to be necessary.

\subsection{Theoretical Framework}
Scholars have examined students’ intention to use GenAI from different theoretical perspectives. On the one hand, each of these theories has its own focus. For example, the Technology Acceptance Model (TAM) focuses on explaining users’ acceptance behaviour towards information technology, while the Expectancy Value Theory (EVT) focuses on how individuals make decisions based on expectancy and values. On the other hand, these theories also involve a certain continuity. For instance, subjective norms in the Theory of Planned Behaviour (TPB) are developed into social influence in the Unified Theory of Acceptance and Use of Technology (UTAUT). Integrating these key theories can provide a comprehensive perspective for analysing students’ intention to use GenAI. In addition, because of the small sample sizes in examining some variables, we will merge these relevant variables and eliminate those without sufficient empirical evidence, in order to ensure the accuracy and representativeness of the results included in the meta-analysis process.

\subsubsection{Theory of Planned Behaviour (TPB)}
TPB was introduced by social psychologist Ajzen (1991)\citep{ajzen1991theory} in the late 1980s, aiming to explain how an individual’s behavioural intention is formed. It is regarded as one of the most influential theories for predicting and explaining human social behaviour nowadays\citep{ajzen2011theory}. In TPB, behavioural intentions are defined as prerequisites for actual actions, and the two are influenced by attitudes, subjective norms, and perceived behavioural control comprehensively, where perceived behavioural control directly influences behaviour to some extent \citep{ajzen2015theory}.

In the context of college students’ GenAI utilisation, behavioural attitudes imply their evaluation judgements (which can be both positive and negative) about the technology utilisation behaviour and its outcomes. Subjective norms relate to students’ expectancy about the support level for GenAI utilisation behaviours from society and important others around them (e.g., teachers, classmates, and family members). In such settings, perceived behavioural control is defined as the student’s subjective assessment of the difficulty in using GenAI.

TPB has been widely used in the field of educational technology, focusing on measuring teachers’ and students’ intention towards using new technologies in education \citep{cheng2016tertiary,chu2016good,puah2022investigating}. Recently, there have also been several studies that have used TPB as a basis for discussing educators’ and students’ perceptions of GenAI. For example, Ivanov et al. \citep{ivanov2024drivers}found that the core variables of the TPB all can well predict the intentions of lecturers and students to use GenAI in higher education institutions, providing empirical evidence to examine the explanatory power of the theory.

\subsubsection{Technology Acceptance Model (TAM) and Unified Theory of Acceptance and Use of Technology (UTAUT)}

TAM is a theoretical model that explains the behavioural mechanisms by which users accept and use new technologies. Davis et al. (1989) were inspired by the Theory of Reasoned Action (TRA) and TPB in constructing the TAM, and based on which the two core conceptions, i.e., perceived usefulness and perceived ease of use, are proposed. The two key factors are mediated by the user’s attitude toward using technology, which affects technology utilisation intention and actual system use (perceived usefulness also directly affects behavioural intention) \citep{davis1996critical}. TAM has received extensive attention from researchers in educational technology, supported by a large body of empirical evidence \citep{al2020investigating,scherer2019technology,wu2017continuance}. However, with iterative technology processes, many researchers have rebuilt the TAM, and successively proposed TAM2 and TAM3 \citep{faqih2015assessing,venkatesh2000theoretical}. Notably, UTAUT and its extended version UTAUT2 were developed on the basis of TAM, which takes moderators into account. The core conceptions in UTAUT2 are the most extensive, including not only performance expectancy, effort expectancy, social influence, and facilitating conditions in UTAUT, but also hedonic motivation, price value, and habit \citep{tamilmani2021extended}.

A number of studies addressing the factors influencing students’ intention to use GenAI have been conducted under TAM, UTAUT, or their extended versions as theoretical frameworks. For example, Rahman et al. \citep{rahman2022examining} utilised expanded TAM and found that perceived usefulness, perceived ease of use, and perceived informativeness positively impacted ChatGPT utilisation attitude, and the attitude had a significant positive effect on students’ intention to use GenAI.

\subsubsection{Expectancy Value Theory (EVT)}
EVT suggests that achieving a certain task is influenced by individual expectancies and value beliefs \citep{backfisch2021variability}. The individual value beliefs towards the task can be subdivided into several parts: utility value, intrinsic value, achievement value, and costs\citep{eccles2002motivational}. Some scholars have analysed students’ intentions to use GenAI through EVT. Sankaran et al. \citep{sankaran2023student} found that perceived value had a significant positive effect on utilisation intention, whereas perceived costs did not significantly impact the intention. There are various benefits of students’ ChatGPT utilisation in higher education, such as instant support, enhanced interactivity and personalised learning experience, but also some concerns. Chan and Zhou \citep{chan2023expectancy} reached similar conclusions.

\section{Research Design}\label{}
Comprehensive Meta-Analysis 3.7 (CMA 3.7) software was employed in this study as a tool to conduct a meta-analysis, providing a comprehensive examination of the factors influencing college students’ behavioural intentions to use GenAI. The research process included literature search and screening, data coding, publication bias analysis, heterogeneity test, effect size calculation and moderating effect test.

\subsection{Literature Search}

Due to the introduction of GenAI still remains relatively new \citep{lambert2023chatgpt}, the research on this issue mainly focuses on the last five years. Therefore, this study mainly collects online literature published in Web of Science, EBSCO, Wanfang Data E-Resources, and China National Knowledge Infrastructure (CNKI) . The literature search strategy was determined as follows: First, TS= (“Generative Artificial Intelligence” OR “GAI” OR “AIGC” OR “AI Generated Content”) AND (“ChatGPT” OR “GPT-4” OR “Alpha Code” OR “GitHub Copilot” OR “Bard”) AND (“Student” OR “Learner ”), with the time range limited from January 2019 to May 2024, the language limited to English, and the databases limited to Web of Science and EBSCO, from which we collected a total of 2,636 articles. Second, TS= (“Generative Artificial Intelligence” OR “ChatGPT”) AND (“Education”), with the time range limited from January 2019 to May 2024, the language limited to Chinese, and the databases limited to Wanfang Data E-Resources and CNKI, from which we collected a total of 1,522 articles.

\begin{figure}
  \centering
  \includegraphics{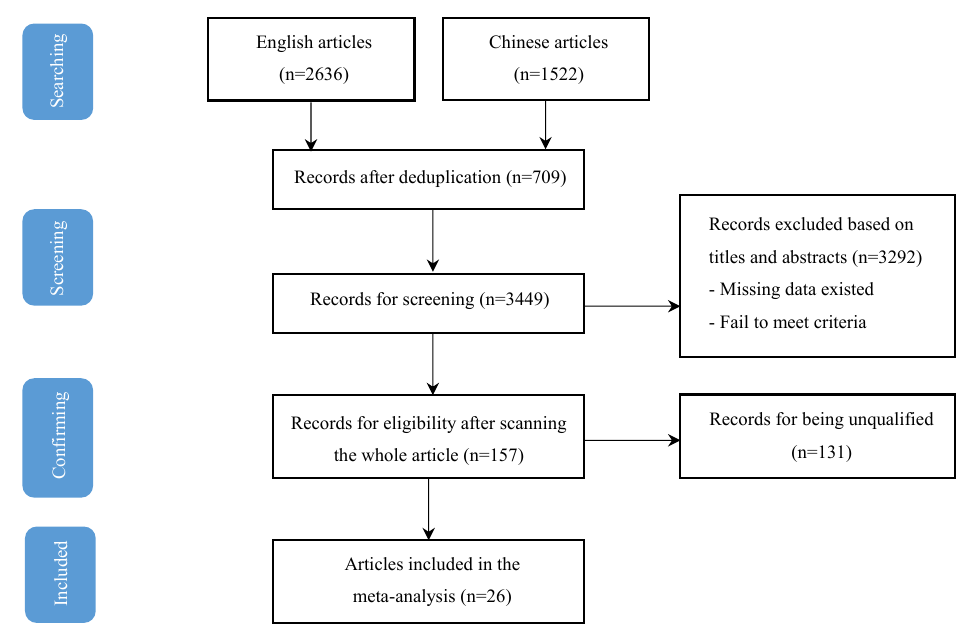}
    \caption{PRISMA Flow Diagram}\label{fig1}
\end{figure}

\subsection{Literature Screening}

After the initial literature search, we conducted a thorough review of the literature included in the meta-analysis with the following criteria: (1) deduplication; (2) focusing on the topic of students’ intention to use GenAI; (3) including quantitative research as sample literature, while excluding qualitative research, theoretical research, and review literature; (4) selecting Chinese or English as the language utilised in sample literature; (5) including the sample size and the correlation coefficient r or any statistical quantities that can be transformed into correlation coefficients in the research data of the sample literature; (6) to ensure independent sample, when same research data being published multiple times, independent variables all are included in the meta-analysis if they are different, while if they are the same, then the literature with a more comprehensive independent variable is retained. The specific screening process is shown in Figure 1. After literature screening, a total of 26 literature samples were included in this meta-analysis, including 24 English-language literature and 2 Chinese-language literature, also 24 journal literature and 2 conference papers. All these GenAI user intention studies focused on college students.

\subsection{Data Coding}
A total of 26 works of literature were included in this study, including 27 studies, two of which were included in the article by Strzelecki et al. \citep{strzelecki2024investigation}. The research regions involved Mainland of China, Hong Kong Special Administrative Region of China, Vietnam, Malaysia, South Korea, Spain, India, and Egypt, etc. After literature screening, the literature included in the meta-analysis was coded for source information (first author, publication time, and literature type), regional category (developing countries/regions, developed countries/regions), sample information (sample size, study object, and gender), and influencing factors and effect sizes. When an article contains multiple independent samples and reports multiple effect sizes, the effect sizes are coded separately.

During effect size coding, we summarised the variables with similar conceptions and deleted the variables with effect sizes below 3, in order to ensure conceptual semantic unification. Then, 9 antecedent variables, including performance expectancy, effort expectancy, social influence, attitude, hedonic motivation, facilitating conditions, perceived behavioural control, perceived cost, and habit, 87 independent study effect sizes and 33,833 sample data were included in this meta-analysis, and defined as shown in Table 2, according to Venkatesh et al. \citep{venkatesh2003user}. In the effect size calculation process, the correlation coefficient r between GenAI utilisation intention and influencing factors was used as the effect size; and if regression coefficients $\beta$ were reported, then they were converted into correlation coefficients according to equation (1), drawing on Peterson and Brown \citep{peterson2005use}.

\begin{center}
\begin{equation}
\label{eq:my_equation} 
r = \beta + 0.05\gamma \quad \text{(when } \beta \geq 0, \gamma = 1; \text{ and when } \beta < 0, \gamma = 0)
\end{equation}
\end{center}




\setcounter{MaxMatrixCols}{20} 
\begin{sidewaystable}[]
\caption{Data Coding (Part)}
\label{tab:my-table}
\resizebox{\columnwidth}{!}{%
\begin{tabular}{lllL{2cm}llL{5cm}L{14cm}}
\hline
\multirow{2}{*}{number} &
  \multicolumn{3}{l}{Source   Information} &
  \multicolumn{2}{l}{Sample   Information} &
  \multirow{2}{*}{Regional   Category} &
  \multirow{2}{*}{\begin{tabular}[c]{@{}l@{}}Influencing   Factors\\    \\ (Correlation   Coefficient /$\beta$-Value)\end{tabular}} \\ \cline{2-6}
 &
  First   Author &
  Publication   Time &
  Literature   Category &
  Sample   Size &
  \begin{tabular}[c]{@{}l@{}}Gender   Ratio\\    \\ (Male)\end{tabular} &
   &
   \\ \hline
1 &
  Cai Qianqian &
  2023 &
  J &
  458 &
  32.10 &
  Mainland of China &
  performance   expectancy (0.49), perceived satisfaction (0.31) \\
2 &
  Cecilia Ka Yuk Chan &
  2023 &
  J &
  405 &
  51.40 &
  Hong Kong Special Administrative Region (SAR) of   China &
  use frequency   (0.339), students' knowledge of AI (0.178), perceived value (0.606), attainment   value (0.587), intrinsic value (0.459), utility value (0.506), perceived cost   (-0.295) \\
3 &
  Cong Doanh Duong &
  2023 &
  J &
  1389 &
  44.20 &
  Vietnam &
  performance   expectancy (0.528), effort expectancy (0.457) \\
4 &
  Behzad Foroughi &
  2023 &
  J &
  406 &
  46.30 &
  Malaysia &
  performance   expectancy (0.207), effort expectancy (0.132), hedonic motivation (0.151), learning   value (0.175), social influence (0.056), facilitating conditions (0.075),   habit (0.045) \\
5 &
  Alexander A. Hernandez &
  2023 &
  C &
  299 &
  27.42 &
  the Philippines &
  perceived   usefulness (0.118), herding (0.106), convenience (0.466), ethical   consideration (0.133), perceived ease of use (0.09), social influence (0.058),   trust (0.061) \\
6 &
  Hyeon Jo &
  2023 &
  J &
  232 &
  41.20 &
  South Korea &
  attitude (0.256),   subjective norms (0.504), trust (0.19), perceived behavioural control (0.074) \\
7 &
  Hyeon Jo &
  2023 &
  J &
  233 &
  41.60 &
  South Korea &
  satisfaction (0.157),   organisational culture (0.156), social influence (0.511), knowledge   acquisition (0.16) \\
8 &
  Greeni Maheshwari &
  2023 &
  J &
  108 &
  21.00 &
  Vietnam &
  perceived   ease of use (0.269), interactivity (-0.604), personalisation (0.443),   perceived usefulness (0.429), trust (-0.057), perceived intelligence (0.604) \\
9 &
  Md. Shahinur Rahman &
  2023 &
  J &
  344 &
  62.20 &
  Bangladesh &
  attitude   (0.832) \\
10 &
  José-María Romero-Rodríguez &
  2023 &
  J &
  400 &
  27.50 &
  Spain &
  performance   expectancy (0.102), hedonic motivation (0.143), cost value (0.069), habit (0.509),   effort expectancy (-0.001), social influence (0.005), facilitating conditions   (0.019) \\
11 &
  Prema Sankaran &
  2023 &
  C &
  200 &
  51.00 &
  India &
  student’s knowledge   (0.23), effort expectancy (0.18), performance expectancy (0.78), intrinsic   value (0.17), perceived cost (0.01) \\
12 &
  Artur Strzelecki &
  2023 &
  J &
  503 &
  53.30 &
  Poland &
  performance   expectancy (0.26), effort expectancy (0.079), social influence (0.093),   hedonic motivation (0.187), cost value (0.083), habit (0.339), creativity   (0.086), facilitating conditions (0.58) \\
13 &
  Chandan Kumar Tiwari &
  2023 &
  J &
  375 &
  46.40 &
  Oman &
  attitude (0.649) \\
14 &
  Min Zo &
  2023 &
  J &
  242 &
  62.40 &
  Mainland of China &
  perceived   usefulness (0.577), attitude (0.85), perceived ease of use (0.689) \\
15 &
  Artur Strzelecki &
  2023 &
  J &
  543 &
  53.04 &
  Poland &
  performance expectancy (0.504), effort expectancy (0.23), social influence (0.192) \\
16 &
  Artur Strzelecki &
  2023 &
  J &
  385 &
  51.17 &
  Egypt &
  performance   expectancy (0.285), effort expectancy (0.013), social influence (0.198) \\
17 &
  Chi Zhang (in Chinese) &
  2023 &
  J &
  190 &
  51.58 &
  Mainland of China &
  perceived   usefulness (0.653), perceived ease of use (0.583), perceived risk (-0.439),   technology anxiety (-0.383), social influence (0.372) \\
18 &
  Anmol Gulati &
  2024 &
  J &
  309 &
  56.30 &
  India &
  performance   expectancy (0.149), effort expectancy (0.146), social influence (0.140),   hedonic motivation (0.138), habit (0.216), system flexibility (0.128), perceived   risk (0.064) \\
19 &
  Tobias Greitemeyer &
  2024 &
  J &
  612 &
  35.62 &
  Austria &
  attitude (0.44),   subjective norms (0.22), perceived behavioural control (0.29) \\
20 &
  Acosta‑Enriquez &
  2024 &
  J &
  499 &
  47.70 &
  Peru &
  perceived   importance (0.193), perceived ease of use (0.204), interest (0.238), positive   feeling (0.418), acceptance (0.441), boredom (-0.145), perceived opportunity   (-0.002), perceived risk (-0.104), perceived usefulness (-0.033), negative   feeling (-0.017) \\
21 &
  Thi Thuy &
  2024 &
  J &
  513 &
  60.04 &
  Vietnam &
  perceived   usefulness (0.255), satisfaction (0.315), knowledge acquisition (0.184), knowledge   share (0.057), knowledge appliance (0.076) \\
22 &
  Naveed Saif &
  2024 &
  J &
  156 &
  80.13 &
  Islamic Pakistan &
  perceived   ease of use (0.041), attitude (0.237) \\
23 &
  Wenjuan Zhu &
  2024 &
  J &
  226 &
  60.62 &
  Mainland of China &
  value expectancy   (0.317), hedonic motivation (0.181), cost value (0.144), social influence (0.217),   ethical consideration (0.188), perceived ease of use (-0.096), facilitating   conditions (0.059), habit (0.058), AI ethical anxiety (-0.087), perceived   ethical risk (-0.021) \\
24 &
  Abu Elnasr &
  2024 &
  J &
  520 &
  54.81 &
  Saudi Arabia &
  social   influence(0.445), performance expectancy(0.398), effort expectancy(0.144), facilitating conditions (-0.204) \\
25 &
  Haodong Chang &
  2024 &
  J &
  303 &
  52.15 &
  Mainland of China &
  attitude (0.278),   subjective norms (0.158), perceived behavioural control (0.455) \\
26 &
  Arif Mahmud &
  2024 &
  J &
  369 &
  63.69 &
  Bangladesh &
  attitude (0.808) \\
27 &
  Jing Zhao (in Chinese) &
  2024 &
  J &
  374 &
  47.00 &
  Mainland of China &
  performance   expectancy (0.34), effort expectancy (0.17), hedonic motivation (0.06), social   influence (0.02),t ask-technology Fit (0.06), creativity (0.19) \\ \hline
\end{tabular}%
}
\end{sidewaystable}

\setcounter{MaxMatrixCols}{20} 
\begin{table}[]
\caption{Summarised Variables and Coded Effect Sizes}
\label{tab:my-table}
\resizebox{\textwidth}{!}{%
\begin{tabular}{L{2cm}L{5cm}L{7cm}llL{1cm}}
\hline
\multirow{2}{*}{Factor} &
  \multirow{2}{*}{Description} &
  \multirow{2}{*}{Definition} &
  \multirow{2}{*}{K} &
  \multicolumn{2}{l}{Correlation Coefficient} 
   
   \\ \cline{5-6}
   &
   &
   &
   &
  Min. &
  Max. 
   \\ \hline
Performance Expectancy &
  performance expectancy, perceived usefulness, learning value, utility   value, value expectancy &
  The   extent to which students believe that using GenAI will help them progress academically &
  19 &
  -0.033 &
  0.703 
   \\
Effort Expectancy &
  effort expectancy, perceived ease of use &
  The   ease with which students use GenAI &
  16 &
  -0.096 &
  0.739 
   \\
Social Influence &
  social influence, subjective norms &
  The   extent to which important others (e.g., family, teachers, friends, etc.)   think students should use GenAI &
  15 &
  0.055 &
  0.561 
   \\
Attitude &
  attitude, trust &
  Students’   overall evaluation and sentiment tendency towards using GenAI &
  10 &
  -0.057 &
  0.900 
   \\
Facilitating Conditions &
  facilitating conditions, convenience, perceived behavioural control &
  The   extent to which students believe that organisational and technical   infrastructure exists to support using GenAI &
  9 &
  -0.204 &
  0.630 
   \\
Hedonic Motivation &
  hedonic motivation, intrinsic value &
  Students’   tendency to use GenAI because of the pleasure that occurs during using GenAI &
  8 &
  0.110 &
  0.459 
   \\
Perceived Cost &
  perceived cost, cost, cost value &
  The   costs (e.g., time, money or other resources, etc.) that students expect to   incur in using GenAI &
  5 &
  -0.295 &
  0.060 
   \\
Habit &
  habit &
  Students’   tendency to use GenAI as a result of previous experience and repetitive usage &
  5 &
  0.095 &
  0.559  \\ \hline
\end{tabular}%
}
\end{table}

\section{Results}\label{}

\subsection{Publication Bias Analysis}
In this study, funnel plots, fail-safe N and Egger's regression test results were used as criteria for analysing publication bias. When the scatter points in the funnel plot are symmetrically distributed on either side of the average effect size, the fail-safe N is not less than 5K+10 (K, number of studies) \citep{rosenthal1979file}, and the p of the Egger's test is greater than 0.05 \cite{egger1997bias}, it indicates a lack of publication bias; conversely, it suggests the presence of publication bias. The funnel plot of this study is shown in Figure 2, where the horizontal axis represents the Fisher’s Z, the vertical axis represents the standard error, and the vertical line represents the combined effect size. From Figure 2, it can be observed that the effect sizes of each antecedent variable exhibit a symmetrical distribution, suggesting a low possibility of publication bias in the data. The results of the fail-safe N and Egger's test are presented in Table 3, where the NFS values for each antecedent variable are all greater than 5K+10, and the p for each antecedent variable in the Egger's test are all greater than 0.05. Therefore, the above tests support that no publication bias exhibited in the effect sizes included in the present study.

\begin{figure}[!h]
  \centering
  \includegraphics[width=.9\textwidth]{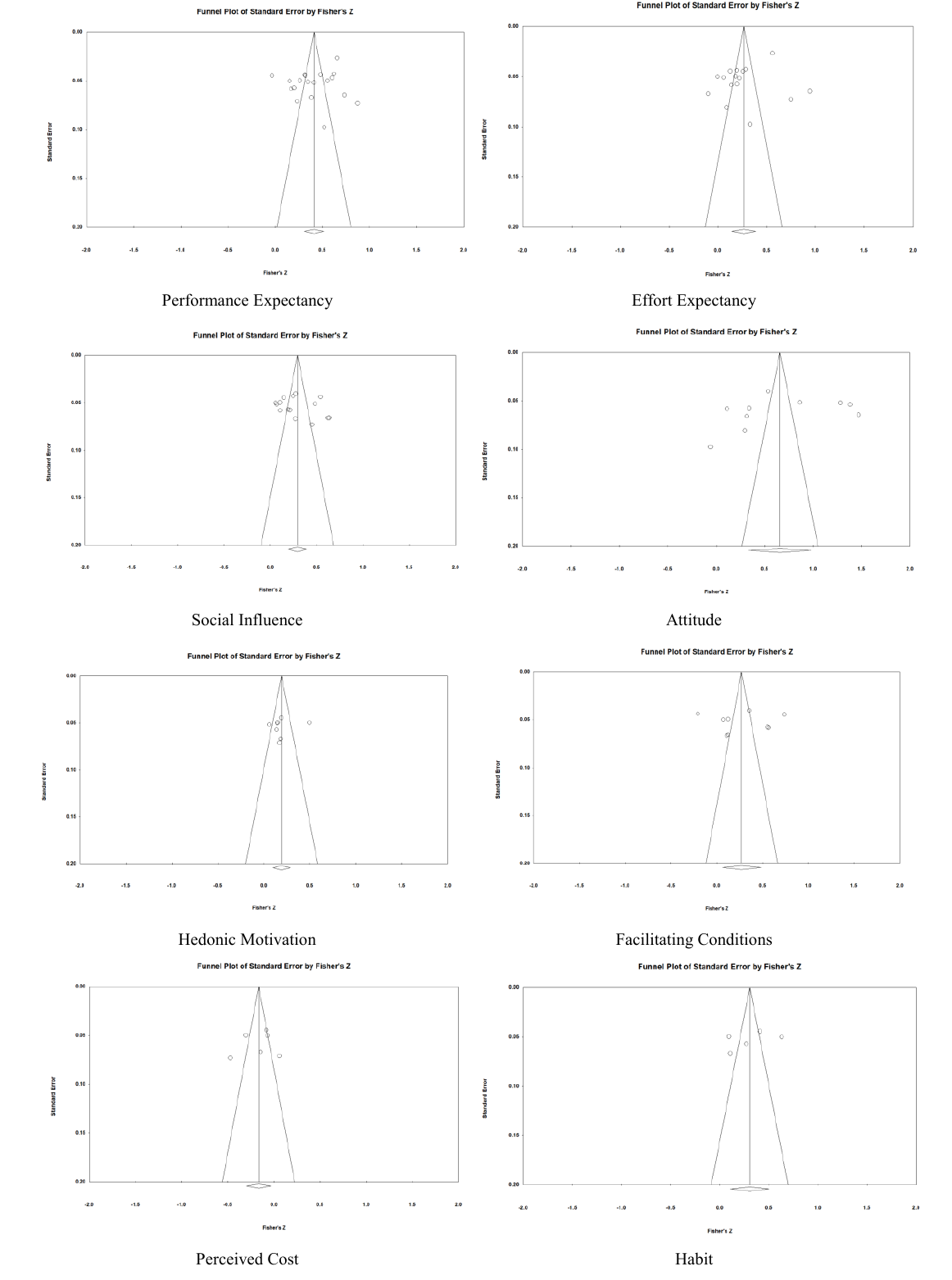}
    \caption{The Funnel Plots of Results of Publication Bias}\label{fig1}
\end{figure}

\begin{table}[]
\caption{Results of Publication Bias Analysis (Egger’s Regression Test)}
\label{tab:my-table}
\fontsize{10pt}{12pt}\selectfont
\begin{tabular}{lllllllll}
\hline
\multirow{3}{*}{Factor} &
  \multirow{3}{*}{K} &
  \multicolumn{2}{l}{Fail-Safe N} &
  \multicolumn{5}{l}{Egger’s Regression Test} \\ \cline{3-9} 
 &
   &
  \multirow{2}{*}{NFS} &
  \multirow{2}{*}{5k+10} &
  \multirow{2}{*}{Intercept} &
  \multicolumn{2}{l}{Confidence Interval (95\%)} &
  \multirow{2}{*}{t-Value} &
  \multirow{2}{*}{p-Value} \\ \cline{6-7}
 &
   &
   &
   &
   &
  Lower Limit &
  \begin{tabular}[c]{@{}l@{}}Upper\\    \\ Litmit\end{tabular} &
   &
   \\ \hline
Performance Expectancy  & 19 & 6224 & 105 & -4.030  & -12.381 & 4.321  & 1.018 & 0.323 \\
Effort Expectancy       & 16 & 1808 & 90  & -4.560  & -13.464 & 4.346  & 1.098 & 0.291 \\
Social Influence        & 15 & 1665 & 85  & 4.937   & -6.964  & 16.838 & 0.896 & 0.396 \\
Attitude                & 10 & 3647 & 60  & -9.720  & -40.183 & 20.743 & 0.376 & 0.483 \\
Facilitating Conditions & 9  & 578  & 55  & -0.027  & -32.087 & 32.033 & 0.002 & 0.998 \\
Hedonic Motivation      & 8  & 209  & 50  & -2.374  & -19.145 & 14.397 & 0.346 & 0.741 \\
Perceived Cost          & 5  & 71   & 35  & -2.670  & -21.636 & 16.296 & 0.391 & 0.716 \\
Habit                   & 5  & 231  & 35  & -12.919 & -60.484 & 34.646 & 0.864 & 0.451 \\ \hline

\end{tabular}%
\end{table}

\subsection{Heterogeneity Test}
The heterogeneity test is used to assess the effect-size variability among different studies. In this study, the Q-statistic and $I^2$ are used to measure the heterogeneity among the included studies. When the significance level of Q is less than 0.05, it indicates the presence of heterogeneity among the studies. $I^2$ reflects the degree of heterogeneity, with boundaries at 25\%, 50\%, and 75\%, representing low, moderate, and high levels of heterogeneity, respectively \citep{higgins2003measuring}. The results of the heterogeneity test determine the mixed-effect model. If heterogeneity exists among studies, a random-effects model is chosen for combining effect size; and if the studies are homogeneous, a fixed-effects model is selected. The results of the tests are presented in Table 4. From Table 4, it can be observed that the Q test results for variables such as performance expectancy, effort expectancy, social influence, attitude, hedonic motivation, facilitating conditions, perceived behavioural control, perceived cost, and habit are significant, and the $I^2$ for each influencing factor are all greater than 75\%, indicating a strong level of heterogeneity. Therefore, a random-effects model is chosen for combining effect sizes.

\begin{table}[]
\caption{Heterogeneity Test for Each Influencing Factor}
\label{tab:my-table}
\fontsize{10pt}{12pt}\selectfont
\begin{tabular}{lllllll}
\hline
\multirow{2}{*}{Factor} & \multirow{2}{*}{K} & \multirow{2}{*}{N} & \multicolumn{3}{l}{Heterogeneity (Q-Test)} & \multirow{2}{*}{Tau-squared} \\ \cline{4-6}
                        &    &      & Q       & P     & $I^2$     &       \\ \hline
Performance Expectancy  & 19 & 7969 & 382.533 & 0.000 & 95.295 & 0.050 \\
Effort Expectancy       & 16 & 6549 & 368.431 & 0.000 & 95.929 & 0.060 \\
Social Influence        & 15 & 5535 & 183.542 & 0.000 & 92.372 & 0.033 \\
Attitude                & 10 & 3040 & 712.479 & 0.000 & 98.737 & 0.265 \\
Facilitating Conditions & 9  & 3501 & 319.327 & 0.000 & 97.495 & 0.102 \\
Hedonic Motivation      & 8  & 2823 & 46.065  & 0.000 & 84.804 & 0.016 \\
Perceived Cost          & 5  & 1734 & 42.232  & 0.000 & 88.161 & 0.024 \\
Habit                   & 5  & 1844 & 73.467  & 0.000 & 94.555 & 0.048 \\ \hline
\end{tabular}%
\end{table}

\subsection{Effect Size Calculation}
Based on the results of the heterogeneity tests as shown in Table 5, a random-effects model was applied to calculate the average effect sizes of each influencing factor on students' intention to use GenAI. According to Cohen's effect size criteria\citep{cohen1988statistical}, when 0.1 ≤ r < 0.3, it indicates a weak correlation; when 0.3 ≤ r < 0.5, it indicates a moderate correlation; when r ≥ 0.5, it indicates a strong correlation, with significance judged based on the p.

From Table 5, it can be observed that perceived cost (r = -0.166) has a weak negative correlation with students’ intention to use GenAI, indicating that as perceived cost increases, students’ intention to use GenAI decreases. Hedonic motivation (r = 0.190), facilitating conditions (r = 0.265), effort expectancy (r = 0.231), social influence (r = 0.280), and habit (r = 0.284) exhibit significant positive correlations with students’ intention to use GenAI. In terms of effect size, habit, social influence, facilitating conditions, effort expectancy, and hedonic motivation show gradually decreasing influence power. Performance expectancy (r = 0.389) is moderately positively correlated with students’ intention to use GenAI, while attitude (r = 0.576) shows a highly significant positive correlation.

Therefore, it can be concluded that performance expectancy (r = 0.389) and attitude (r = 0.576) are key variables influencing students’ intention to use GenAI.

\begin{table}[]
\caption{Effect Size Calculation by Influencing Factors}
\label{tab:my-table}
\fontsize{10pt}{12pt}\selectfont
\begin{tabular}{lllllll}
\hline
\multirow{2}{*}{Factor} & \multirow{2}{*}{Model} & \multirow{2}{*}{R’s Merge} & \multicolumn{2}{l}{Correlation (95\% CI)} & \multicolumn{2}{l}{Two-Tailed Test} \\ \cline{4-7} 
                        &                      &        & LL     & UL     & Z-Value & p-Value \\ \hline
Performance Expectancy  & Random-effects Model & 0.389  & 0.298  & 0.473  & 7.789   & 0.000   \\
Effort Expectancy       & Random-effects Model & 0.259  & 0.141  & 0.370  & 4.219   & 0.000   \\
Social Influence        & Random-effects Model & 0.284  & 0.193  & 0.370  & 5.931   & 0.000   \\
Attitude                & Random-effects Model & 0.576  & 0.323  & 0.753  & 4.003   & 0.000   \\
Facilitating Conditions & Random-effects Model & 0.265  & 0.060  & 0.449  & 2.514   & 0.012   \\
Hedonic Motivation      & Random-effects Model & 0.190  & 0.096  & 0.281  & 3.395   & 0.000   \\
Perceived Cost          & Random-effects Model & -0.166 & -0.292 & -0.034 & -02.459 & 0.014   \\
Habit                   & Random-effects Model & 0.296  & 0.107  & 0.465  & 3.018   & 0.003   \\ \hline
\end{tabular}%
\end{table}

\subsection{Moderating Effect Analysis}
When the moderator variable is the categorical variable, subgroup analysis should be conducted to perform subgroup effect size difference tests based on calculating the effect sizes of each subgroup. 

\subsubsection{Regional Category}
In this study, we followed the classification criteria proposed by the International Monetary Fund. Specifically, Hong Kong Special Administrative Region (SAR) of China, South Korea, Spain, and Austria were classified as developed countries (regions), while Mainland of China, India, Egypt, Malaysia, the Philippines, Poland, Bangladesh, Oman, Pakistan, Saudi Arabia, Vietnam, and Peru were categorised as developing countries (regions). According to Table 6, the relationship between effort expectancy (QB=12.238, p<0.05) and habit (QB=18.293, p<0.05) with students’ behavioural intention to use GenAI is moderated by locational factors. Compared to developed countries (regions), in developing countries (regions), students’ behavioural intention to use GenAI is more likely to be influenced by effort expectancy (r=0.276, p<0.05) and less influenced by habit (r=0.221, p<0.05). Locational factors do not moderate the impact of performance expectancy (QB=0.088, p>0.05), social influence (QB=0.932, p>0.05), attitude (QB=1.466, p>0.05), hedonic motivation (QB=0.944, p>0.05), facilitating conditions (QB=0.469, p>0.05), and perceived cost (QB=0.036, p>0.05) on students’ intention to use GenAI.

\begin{table}[]
\caption{Results of the Moderation Effect Analysis by Regional Category}
\label{tab:my-table}
\resizebox{\textwidth}{!}{%
\begin{tabular}{llllllllllll}
\hline
\multirow{2}{*}{Factor} &
  \multirow{2}{*}{Subgroup} &
  \multirow{2}{*}{K} &
  \multirow{2}{*}{N} &
  \multirow{2}{*}{R’s Merge} &
  \multicolumn{2}{l}{Correlation (95\% CI)} &
  \multicolumn{2}{l}{Two-Tailed Test} &
  \multicolumn{3}{l}{Heterogeneity Analysis} \\ \cline{6-12} 
 &
   &
   &
   &
   &
  LL &
  UL &
  Z-Value &
  P-Value &
  QB(Total between Q) &
  DF &
  P \\ \hline
\multirow{2}{*}{Performance Expectancy} &
  Developed &
  2 &
  805 &
  0.341 &
  -0.041 &
  0.636 &
  1.758 &
  0.079 &
  \multirow{2}{*}{0.088} &
  \multirow{2}{*}{1} &
  \multirow{2}{*}{0.767} \\
 &
  Developing &
  17 &
  7164 &
  0.395 &
  0.298 &
  0.484 &
  7.423 &
  0.000 &
   &
   &
   \\
\multirow{2}{*}{Effort Expectancy} &
  Developed &
  1 &
  400 &
  -0.001 &
  -0.099 &
  0.097 &
  -0.020 &
  0.984 &
  \multirow{2}{*}{12.238} &
  \multirow{2}{*}{1} &
  \multirow{2}{*}{0.000} \\
 &
  Developing &
  15 &
  6149 &
  0.276 &
  0.157 &
  0.388 &
  4.430 &
  0.000 &
   &
   &
   \\
\multirow{2}{*}{Social Influence} &
  Developed &
  4 &
  1477 &
  0.375 &
  0.133 &
  0.575 &
  2.962 &
  0.003 &
  \multirow{2}{*}{0.932} &
  \multirow{2}{*}{1} &
  \multirow{2}{*}{0.334} \\
 &
  Developing &
  11 &
  4058 &
  0.251 &
  0.154 &
  0.343 &
  4.962 &
  0.000 &
   &
   &
   \\
\multirow{2}{*}{Attitude} &
  Developed &
  2 &
  844 &
  0.407 &
  0.214 &
  0.570 &
  3.937 &
  0.000 &
  \multirow{2}{*}{1.466} &
  \multirow{2}{*}{1} &
  \multirow{2}{*}{0.226} \\
 &
  Developing &
  8 &
  2196 &
  0.614 &
  0.302 &
  0.8 &
  3.475 &
  0.001 &
   &
   &
   \\
\multirow{2}{*}{Facilitating Conditions} &
  Developed &
  3 &
  1244 &
  0.184 &
  -0.007 &
  0.361 &
  1.886 &
  0.059 &
  \multirow{2}{*}{0.469} &
  \multirow{2}{*}{1} &
  \multirow{2}{*}{0.494} \\
 &
  Developing &
  6 &
  2257 &
  0.306 &
  -0.004 &
  0.562 &
  1.938 &
  0.053 &
   &
   &
   \\
\multirow{2}{*}{Hedonic Motivation} &
  Developed &
  2 &
  805 &
  0.310 &
  -0.025 &
  0.582 &
  1.818 &
  0.069 &
  \multirow{2}{*}{0.944} &
  \multirow{2}{*}{1} &
  \multirow{2}{*}{0.331} \\
 &
  Developing &
  6 &
  2018 &
  0.147 &
  0.103 &
  0.189 &
  6.604 &
  0.000 &
   &
   &
   \\
\multirow{2}{*}{Perceived Cost} &
  Developed &
  2 &
  805 &
  -0.185 &
  -0.394 &
  0.044 &
  -1.589 &
  0.112 &
  \multirow{2}{*}{0.036} &
  \multirow{2}{*}{1} &
  \multirow{2}{*}{0.850} \\
 &
  Developing &
  3 &
  929 &
  -0.156 &
  -0.338 &
  0.036 &
  -1.592 &
  0.111 &
   &
   &
   \\
\multirow{2}{*}{Habit} &
  Developed &
  1 &
  400 &
  0.559 &
  0.488 &
  0.623 &
  12.580 &
  0.000 &
  \multirow{2}{*}{18.293} &
  \multirow{2}{*}{1} &
  \multirow{2}{*}{0.000} \\
 &
  Developing &
  4 &
  1444 &
  0.221 &
  0.066 &
  0.365 &
  2.775 &
  0.006 &
   &
   &
   \\ \hline
\end{tabular}%
}
\end{table}

\subsubsection{Gender}
In this study, the male ratio in each sample in the included research was encoded as the continuous variable. With gender as the moderating variable, meta-regression analysis was conducted to examine whether the relationship between each influencing factor and students’ intention to use GenAI is significantly influenced by gender. As shown in Table 7, the results indicate that when gender works as the moderating variable, except for attitude (p<0.05), there is no significant moderating effect on performance expectancy, effort expectancy, social influence, hedonic motivation, facilitating conditions, perceived cost, and habit (p>0.05). Therefore, gender significantly moderates the relationship between attitude and students’ intention to use GenAI. As the male ratio increases, the correlation coefficient between attitude and behavioural intention also increases.

\begin{table}[]
\caption{Results of the Moderation Effect Test by Gender}
\label{tab:my-table}
\resizebox{\textwidth}{!}{%
\begin{tabular}{lllllll}
\hline
Factor & Regression   Coefficient & Standard Error & LL & UL & Z-Value & Two-Tailed   Significance \\ \hline
Performance Expectancy  & 0.003  & 0.005 & -0.006 & 0.012 & 0.66  & 0.509 \\
Effort Expectancy       & 0.002  & 0.005 & -0.008 & 0.011 & 0.33  & 0.739 \\
Social Influence        & 0.004  & 0.005 & -0.006 & 0.014 & 0.77  & 0.440 \\
Attitude                & 0.018  & 0.008 & 0.003  & 0.032 & 2.30  & 0.022 \\
Facilitating Conditions & -0.003 & 0.010 & -0.023 & 0.017 & -0.29 & 0.776 \\
Hedonic Motivation      & 0.002  & 0.006 & -0.009 & 0.013 & 0.43  & 0.669 \\
Perceived Cost          & -0.004 & 0.007 & -0.180 & 0.010 & -0.53 & 0.595 \\
Habit                   & -0.013 & 0.007 & -0.027 & 0.002 & -1.70 & 0.089 \\ \hline
\end{tabular}%
}
\end{table}

\section{Discussion and Conclusion}\label{}

\subsection{Discussion}

\subsubsection{General Discussion}
Understanding the influencing factors and underlying mechanisms of students’ behavioural intention to use GenAI is an urgent topic. In this study, we integrated key variables from existing theories (e,g., TPB, TAM, UTAUT, EVT), applied meta-analysis to examine the impact on students’ intention to use GenAI, determined the effect size of these influences, and conducted moderation effect analyses around regional category and gender. Our research provides important insights into the influencing factors of students’ intention to use GenAI. The factors analysed in the study all exhibit significant correlations with students’ intention to use GenAI. Furthermore, the study revealed that performance expectancy (r=0.389) and attitude (r=0.576) show relatively strong correlations, indicating that they are key variables influencing students’ behavioural intention to use GenAI.

Compared to other information technologies, GenAI tools represented by ChatGPT are more accessible. Students can easily access ChatGPT online and obtain assistance from ChatBot with minimal cost \citep{adeshola2023opportunities}. A respondent, according to the study by Menon and Shilpa \citep{menon2023chatting}, even stated, “It is a straightforward application. Even a 5-year-old child can use it with ease. There is no technical support that is required.” This suggests that objective constraints, such as basic infrastructure, technological devices, and economic conditions, have been reduced for students. This may explain why students’ behavioural intention to use GenAI largely depends on their subjective perceptions rather than objective conditions. Therefore, performance expectancy and attitude work as the most critical influencing factors, while the importance of perceived cost (r=-0.166) decreases.

Attitude is the most crucial influencing factor on students’ behavioural intention to use GenAI. Several studies have confirmed that attitude significantly predicts students’ intention to use GenAI \citep{mahmud2024adoption, saif2024chat}, regardless of their digital capacity \citep{chang2024research}. Through meta-analysis, we have provided stronger evidence for the role of attitude in shaping students’ behavioural intention to use GenAI.

Performance expectancy is the second most important influencing factor, which is consistent with previous studies \citep{strzelecki2024students,strzelecki2023use}. It is noteworthy that both existing studies found that habit’s influence surpassed performance expectancy, ranking first. However, this meta-analysis study challenges this assertion. We discovered that while habit (r=0.284) also exhibits a significant positive correlation with students’ behavioural intention to use GenAI, its impact is relatively weaker. Additionally, this study also revealed that the impact of performance expectancy is stronger than that of effort expectancy (r=0.231), supporting the findings of Zhu et al. \citep{zhu2024could}. However, in the meta-analysis study, we did not conduct further analysis on the influence of effort expectancy on performance expectancy. In fact, in the study by Duong et al. \citep{duong2023applying}, effort expectancy can influence students’ behavioural intention to use GenAI through performance expectancy as a mediator.

In addition, the study found that social influence (r=0.280), facilitating conditions (r=0.265), and hedonic motivation (r=0.190) exhibit significant positive correlations with students’ behavioural intention to use GenAI. Although these three variables all originate from UTAUT2, we also integrated subjective norms and perceived behavioural control from TPB in constructing the analytical framework, as well as intrinsic value from EVT. This meta-analysis study supports Jo’s assertion \citep{jo2023decoding}  that subjective norms significantly influence students’ behavioural intention to use ChatGPT but contrasts sharply with their finding that perceived behavioural control does not significantly impact the utilisation intention. Regarding hedonic motivation, the findings of this study align with Strzelecki \citep{strzelecki2023use}, providing evidence for Sankaran et al.’s \citep{sankaran2023student} judgment on the impact of intrinsic value.

\subsubsection{Discussion on Regional Category}

We not only explored the influencing factors of students’ behavioural intention to use GenAI but also identified heterogeneity in two key variables across different regional categories, which could potentially impact our research findings.
Firstly, our findings indicate that, compared to developed countries (regions), students from developing countries (regions) are more likely to be influenced by effort expectancy (r=0.276, p<0.05) in their intention to use GenAI. It could be explained that students in developing countries (regions) may be equipped with weaker information technology skills \citep{skryabin2015ict,vargas2023ict}, and once they perceive the accessibility of GenAI, exemplified by ChatGPT, their intention to use the technology then strengthens. Consistent with Abdaljaleel et al. \citep{abdaljaleel2024multinational}, we demonstrate that the perception of the ease of technology use significantly impacts the GenAI utilisation intention of students in developing countries (regions) to use GenAI. However, it is clear that more research directly comparing students’ behavioural intention to use GenAI across different countries (regions) is needed to confirm this conclusion.

Furthermore, compared to developed countries (regions), students in developing countries (regions) are less likely to be influenced by habit (r=0.221, p<0.05) on their intention to use GenAI. A possible reason for this could be that students in developing countries (regions) have had a late exposure to GenAI, which has not yet led to the habituation. However, as behaviour becomes more mechanical, the importance of intention in determining behaviour decreases. This suggests that when users’ use of an information system becomes habitual, their intention may no longer be the main driver of continued usage \citep{limayem2007habit}. Due to the lack of studies in the meta-analysis examining the differences between intention and habit in the utilisation behaviour of GenAI, we cannot make comparisons. Apparently, future research should make efforts to clarify their effects.

\subsubsection{Discussion on Gender as a Factor}

Our findings indicate that gender only moderates students’ attitudes towards using GenAI, hardly affecting other variables. In sharp contrast to the study by Arthur et al. \citep{arthur2024predictors}, our research does not support the notion that gender plays a moderating role in the relationship between facilitating conditions and students’ behavioural intention to use GenAI. Consistent with the study by Strzelecki et al. \citep{strzelecki2023use}, our research also found that gender has rarely impact on the role of habit. Additionally, our meta-analysis study resolves the debate in existing research \citep{strzelecki2024students} regarding whether gender moderates major variables of the UTAUT model concerning students’ behavioural intention to use GenAI. We found that the higher the proportion of males, the higher the correlation coefficient between attitude and intention to use GenAI. A possible explanation is that based on societal expectations of gender roles, males are generally perceived to be more willing to take risks in decision-making \citep{harris2006gender}, and such a tendency makes them more open-minded in trying out new technologies.

\subsection{Theoretical Implications}
This study integrates classic theories concerning user acceptance of new technologies, including TPB, TAM, UTAUT2, and EVT. We selected core variables based on existing research and formed a comprehensive theoretical framework to analyse students’ behavioural intention to use GenAI. For instance, we utilised perceived cost that combines discussions on price value from UTAUT2 and costs from EVT. Also, we employed performance expectancy from UTAUT2 to interiorise such as perceived usefulness from TAM and utility value from EVT, etc. Such a comprehensive theoretical framework provides new insights into exploring the multidimensional factors that drive or hinder students’ behavioural intention to use GenAI. Furthermore, by utilising the results of our meta-analysis study, we addressed debates in existing research regarding the effect size of different variables, thereby providing empirical support for establishing a research consensus on this topic.

\subsection{Practical Implications}

This study’s practical implication lies in overcoming the limitations of individual research by integrating existing research data, providing more universal conclusions for various stakeholders such as technology developers, school decision-makers, and educators.

Firstly, we found that the operation process of GenAI is relatively user-friendly, reducing the role of effort expectancy while highlighting the importance of attitude and performance expectancy. Therefore, for technology developers, it is crucial to curb the generation of false information through stricter content verification mechanisms, enhance privacy protection for data security, and build trust and positive attitude among students. Additionally, technology developers need to understand the demands of users to ensure that technological solutions address real issues and enhance the adaptability of GenAI to users through improved interface design, personalised settings, and feature guidance services to increase students’ expectations of technological performance.

Secondly, school decision-makers need to encourage the formulation of corresponding guidelines and principles within school settings, establish institutional norms through consensus-building, and leverage social influence at the organisational level. For example, the GenAI technology usage guidelines provided to students by Harvard University (https://provost.harvard.edu/guidelines-using-chatgpt-and-other-generative-ai-tools-harvard) serve as a significant reference for other schools to develop similar regulations.

Lastly, teachers can consider integrating GenAI into their teaching design and guide students in using GenAI properly. On the one hand, technologies like ChatGPT have the potential to create gamified learning environments \citep{fulcini2023chatgpt}. Teachers can use GenAI technology to assist students in gamified learning, stimulate hedonic motivation, optimise student learning experiences, and improve their academic emotions. On the other hand, as important others in students’ lives, teachers should help students form appropriate GenAI technology use habits. Additionally, teachers should remain sensitive to gender differences in students’ behavioural intntion to use GenAI, encouraging female students to adopt an open attitude towards new technologies.

\subsection{Limitations and Directions}

This study also has some limitations that need to be addressed in future research.

Firstly, the number of sample references is limited, and the heterogeneity of samples in terms of different educational stages is not sufficiently diverse. Although many scholars have discussed the significances and limitations of applying GenAI into education, empirical literature specifically related to factors influencing students’ behavioural intention to use GenAI remains very limited, which could potentially affect the results of data analysis. Additionally, the study primarily focuses on students in higher education, with a lack of attention to students in basic education stages, which needs to be addressed in future research.

Secondly, the meta-analysis only includes quantitative studies on factors influencing the behavioural intention to use GenAI, while exploring these factors through qualitative research could provide a richer perspective. Therefore, in future research, a qualitative meta-analysis of factors influencing the behavioural intention to use GenAI should be conducted to comprehensively understand the multidimensional factors influencing the behavioural intention to use GenAI.

Lastly, some moderating variables need to be further explored. This study only analysed the impact of regional category and gender as moderating variables on the antecedent variables, while the analysis of potential moderating variables such as discipline category and urban-rural differences was lacking.

Despite above limitations of this study existed, it has laid a solid foundation for future research. Subsequent studies can provide new insights by expanding sample sizes, optimising data processing methods, and adding additional moderating variables.


\end{document}